\documentclass[english,10pt,aps,prd,a4paper,preprintnumbers,floatfix,nofootinbib,showpacs,superscriptaddress]{revtex4-1}
\usepackage[english]{babel}
\usepackage{amsmath}
\usepackage{graphicx}
\usepackage{color}
\usepackage{mathrsfs}   
\usepackage{amssymb}
\usepackage{hyperref}
\usepackage[normalem]{ulem} 
\usepackage{dcolumn}
\usepackage{bm}
\usepackage{graphicx}
\usepackage[sort&compress]{natbib}
\usepackage{epstopdf}
\newcommand {\be} {\begin{equation}}
\newcommand {\ee} {\end{equation}}

\definecolor{greenLinks}{rgb}{0, 0.6, 0} 
\definecolor{blueLinks}{rgb}{0, 0, 0.6}
\definecolor{redLinks}{rgb}{0.6, 0, 0}
\definecolor{tempText}{rgb}{0.55, 0.10,0.67}
\definecolor{eprintLinks}{rgb}{0.4, 0.4, 0.4}
\definecolor{journalLinks}{rgb}{0.6, 0, 0}

\newcommand{\MYhref}[3][redLinks]{\href{#2}{\color{#1}{#3}}}%

\def\vev#1{\left\langle #1\right\rangle}

\usepackage{float}

\def\vev#1{\left\langle #1\right\rangle}

\def\21{$\mathrm{SU(2)_L \otimes U(1)_Y}$ }
\def\31{$\mathrm{SU(3)_c \otimes U(1)_Q}$ }
\def\SM{$\mathrm{SU(3)_c \otimes SU(2)_L \otimes U(1)_Y}$ }

\def\3211{$\mathrm{SU(3) \otimes SU(2)_L \otimes U(1)_R \otimes U(1)_{B-L}}$ }
\def\321{$\mathrm{SU(3) \otimes SU(2) \otimes U(1)}$ }
\def\422{$\mathrm{SU(4) \otimes SU(2) \otimes SU(2)_R}$ }
\newcommand {\ignore}[1]{}

\newcommand{\sm}{{Standard Model }}

\def\vev#1{\left\langle #1\right\rangle}

\def\SM{$\mathrm{ SU(3)_C \otimes SU(2)_L \otimes U(1)_Y }$ }

\newcommand{\AddrAHEP}{%
  AHEP Group, Institut de F\'{i}sica Corpuscular --
  C.S.I.C./Universitat de Val\`{e}ncia, Parc Cient\'ific de Paterna.\\
 C/ Catedr\'atico Jos\'e Beltr\'an, 2 E-46980 Paterna (Valencia) - SPAIN}

\begin{document}

\title{Seesaw roadmap to neutrino mass and dark matter}

\author{Salvador Centelles Chuli\'{a}}\email{salcen@ific.uv.es}
\affiliation{\AddrAHEP}
\author{Rahul Srivastava}\email{rahulsri@ific.uv.es}
\affiliation{\AddrAHEP}
\author{Jos\'{e} W. F. Valle}\email{valle@ific.uv.es}
\affiliation{\AddrAHEP}

\begin{abstract}
   \vspace{1cm}

   We describe the many pathways to generate Majorana and Dirac
   neutrino mass through generalized dimension-5 operators {\sl a la
     Weinberg}. 
   The presence of new scalars beyond the \sm Higgs doublet implies
   new possible field contractions, which are required in the case of
   Dirac neutrinos.
   We also notice that, in the Dirac neutrino case, the extra
   symmetries needed to ensure the Dirac nature of neutrinos can also
   be made responsible for stability of dark matter.

\end{abstract}

\maketitle


\section{Introduction}
\label{sec:introduction}

The discovery of neutrino
oscillations~\cite{Kajita:2016cak,McDonald:2016ixn} represents a
milestone in particle physics, with far--reaching implications.
Indeed, the existence of neutrino mass provides a fundamental guide
for the nature of the new physics that may lie ``beyond the
desert''~\cite{Valle:2015pba}.
Given their charge and color neutrality, massive neutrinos are
generally expected to be Majorana~\cite{Schechter:1980gr} irrespective
of the nature of the mechanism engendering their mass.
However, the characteristic signature~\cite{Schechter:1981bd} of
Majorana neutrinos, namely, the observation of neutrinoless double
beta decay ($0 \nu 2 \beta$), has so far remained
elusive~\cite{Agostini:2017iyd,Albert:2014awa,Alduino:2017ehq}.

In this paper we sketch the landscape of theories where neutrino mass
arises at the dimension-5 operator level within a broader perspective.
As an introduction, we revisit the simplest case of Majorana neutrinos
where, as shown by Weinberg~\cite{weinberg:1980bf}, this operator is
unique if only \sm scalar fields are considered.
Its simplest high energy completion requires new messenger particles.
For example, the tree-level exchange of heavy singlet fermions induces
small neutrino masses through what is now generically called ``type-I
seesaw
mechanism''~\cite{Minkowski:1977sc,GellMann:1980vs,Yanagida:1979as,mohapatra:1980ia,Schechter:1980gr,Schechter:1982cv}.
The heavy messengers can also be extra scalar Higgs bosons, such as
scalar triplets in the ``type-II seesaw''
realization~\cite{Schechter:1980gr}.
The resulting theoretical pathways to Majorana seesaw were nicely
synthesized in Ref.~\cite{ma:1998dn}.

The purpose of this paper is to provide a wider roadmap for tree
level neutrino mass generation within Weinberg's approach.
In doing so we encounter yet unexplored Majorana seesaw
realizations. In addition, we discuss the possibilities for the case
of Dirac seesaw~\footnote{Dirac neutrinos with seesaw-induced masses
  may be introduced at the phenomenological
  level~\cite{Gu:2006dc,Gu:2007mc}. However, consistent UV-complete
  scenarios require the implementation of symmetries protecting
  ``Diracness''.}.
We note that Dirac neutrinos can indeed arise in many symmetry--based
scenarios for generating neutrino mass in generalized versions of the
seesaw~mechanism~\cite{Valle:2016kyz,Bonilla:2016zef,Chulia:2016ngi,Reig:2016ewy,Bonilla:2017ekt,CentellesChulia:2017koy}~\footnote{Likewise,
  Dirac neutrinos naturally emerge in schemes based upon extra
  dimensions~\cite{Chen:2015jta,Addazi:2016xuh}.}.
Motivated by the growing interest in theories with Dirac neutrinos 
\cite{Abbas:2013uqh, Ma:2014qra, Okada:2014vla, Ma:2015rxx,
  Ma:2015raa, Ma:2015mjd, Chen:2015jta, Valle:2016kyz, 
  Bonilla:2016zef, Bonilla:2016diq, Chulia:2016ngi, Chulia:2016giq,
  Reig:2016ewy, Ma:2016mwh, Wang:2016lve, Borah:2016zbd,
  Addazi:2016xuh, Abbas:2016qbl, Heeck:2013rpa, Yao:2017vtm,
  Ma:2017kgb,CentellesChulia:2017koy, Borah:2017leo, Wang:2017mcy,
  Bonilla:2017ekt, Hirsch:2017col, Srivastava:2017sno, Borah:2017dmk,Yao:2018ekp},
we generalize Weinberg's approach to this case.
We sketch the architecture of neutrino mass generation through the
seesaw mechanism when the neutrino mass is of the Dirac
type.
The methodology is the same as in the Majorana case, but there are many
more possibilities, even at the lowest, dimension-5 level.
The reason is that more Higgs multiplets, beyond that of the \sm are
necessarily required, leading to many possible field contractions.
Since in order to obtain Dirac neutrinos we need extra symmetries
beyond those of the Standard Model, we also examine the possibility to
utilize them to stabilize dark matter as well.

The plan of this paper is as follows. In Section
\ref{sec:majorana-neutrinos} we revisit the well known case of
Majorana neutrino mass generation through the Weinberg operator and
its possible seesaw completions.
In addition, we discuss new ways to obtain Majorana neutrino mass
through ``generalized Weinberg'' operators involving new scalar fields.
These can arise in contexts where the standard Weinberg operator is
forbidden by symmetry. 
We also briefly discuss some of their possible seesaw realizations. In
Section \ref{sec:dirac-neutrinos} we move on to the case of Dirac
neutrinos. In this case, with \sm fields only, there is no dimension-5
possibility for Dirac neutrinos. 
Within the minimal \sm Higgs sector, seesaw Dirac neutrino masses can
only be realized at the dimension-6 level or higher, as shown in
Sec.~\ref{dweinberg}.
In Section \ref{dgweinberg} we discuss the generation of Dirac
neutrino mass from generalized Weinberg operators which involve other
scalars beyond the \sm Higgs doublet. We also discuss the various
possibilities for Dirac seesaw completion of such generalized Weinberg
operators.
In order for neutrinos to be Dirac particles, an additional symmetry
is required to protect its Dirac nature. In Section \ref{sec:DM
  sector} we show that, quite generally, such symmetries can be used
to stabilize dark matter~\cite{Chulia:2016ngi}. 


\section{Majorana neutrinos}
\label{sec:majorana-neutrinos}


The aim of this section is to provide a brief introduction to seesaw
constructs and to set up our notations.
We start our task by summarizing the procedure to induce neutrino mass
at the operator level, for the case of Majorana neutrino masses. First
we take the simplest case where the theory only contains \sm fields
and then generalize to the case where new fields are present.


\subsection{Only  Standard Model fields}
\label{weinberg}

It is a well known fact that, for the case of Majorana neutrinos, if only
\sm fields are present, there is a unique dimension-5 operator that
gives rise to neutrino masses, the well-known Weinberg
operator~\cite{weinberg:1980bf},
\be
\frac{1}{\Lambda}  \, \bar{L}^c \otimes \Phi \otimes \Phi \otimes L
\label{weinberg}
\ee  
where $L$ and $\Phi$ denote the lepton and Higgs doublets, and
$\Lambda$ represents the cutoff scale. 
Above the cutoff scale the Ultra-Violet (UV) complete theory is at
play, involving new ``messenger'' fields, whose masses lie close to
the scale $\Lambda$.

\begin{figure}[!h]
 \centering
  \includegraphics[width=0.3\textwidth]{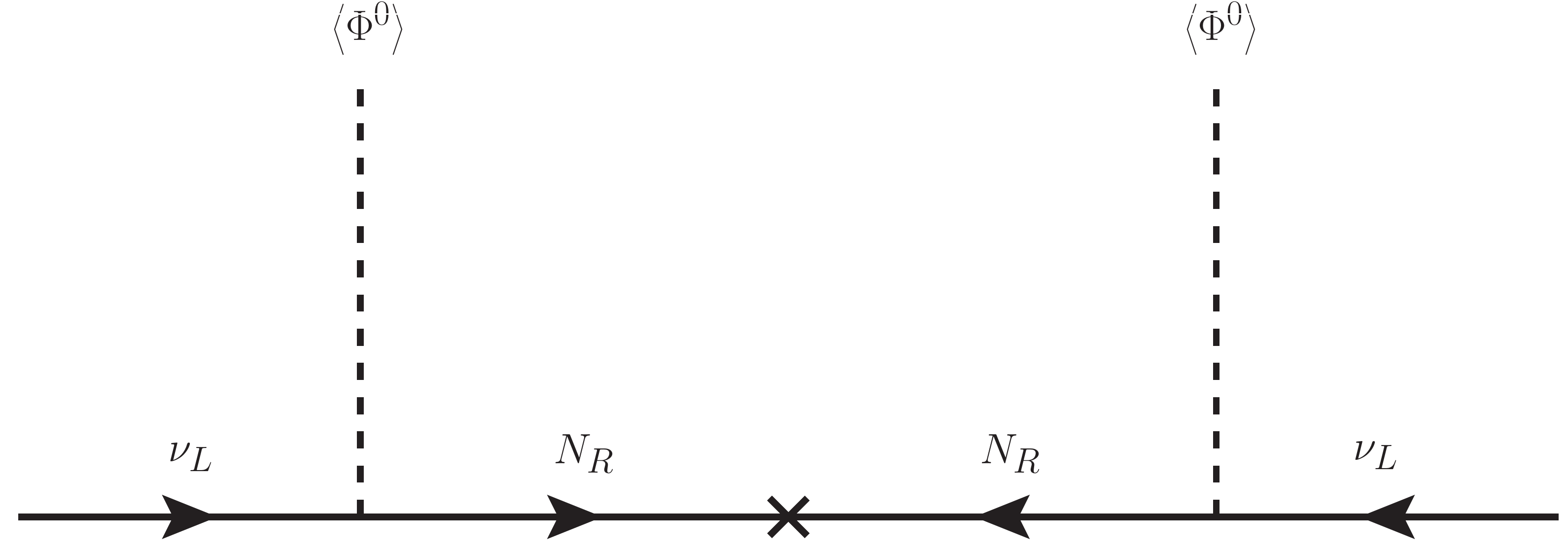}~~~~~~
 \includegraphics[width=0.2\textwidth]{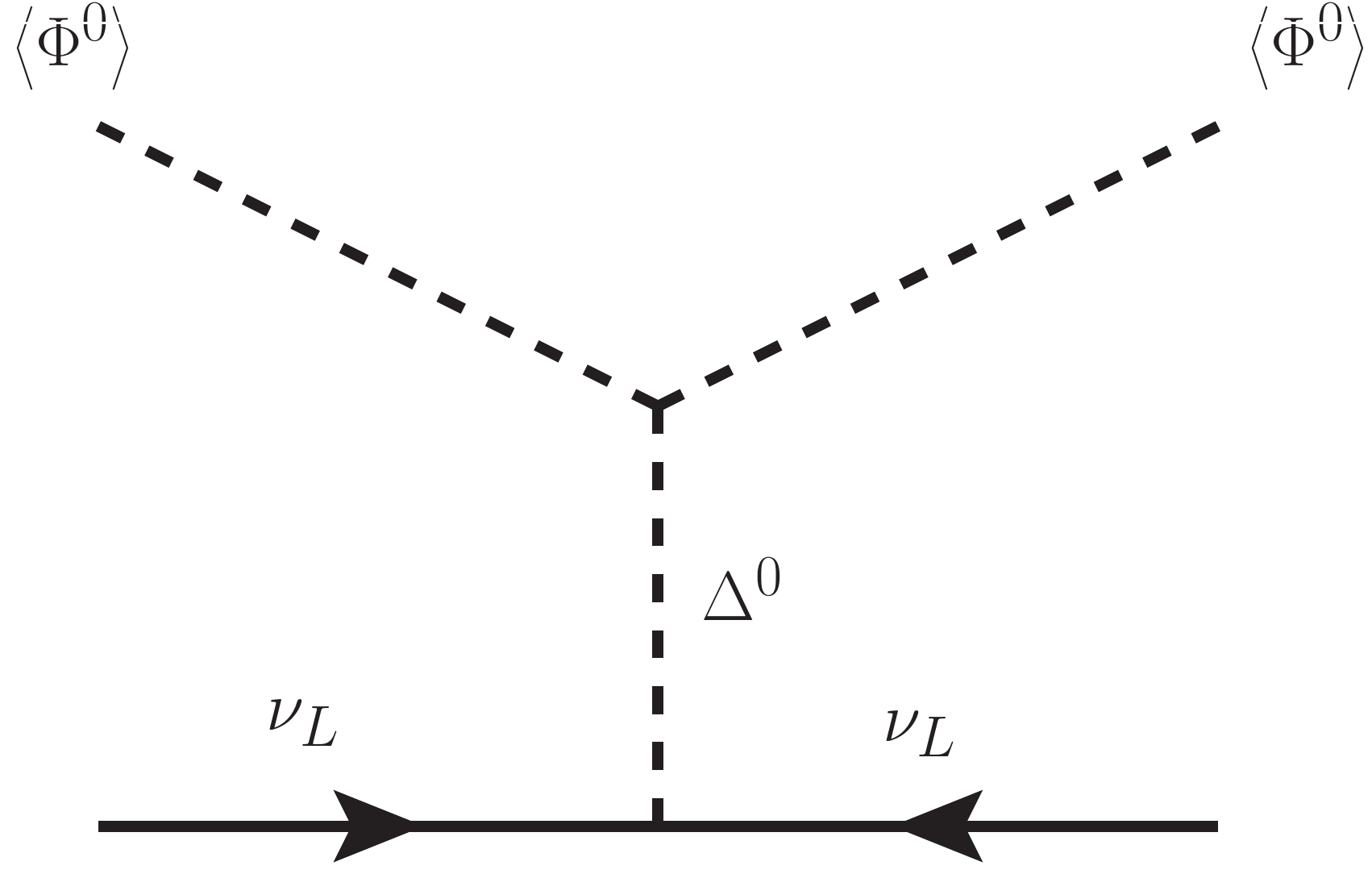}~~~~~~
 \includegraphics[width=0.3\textwidth]{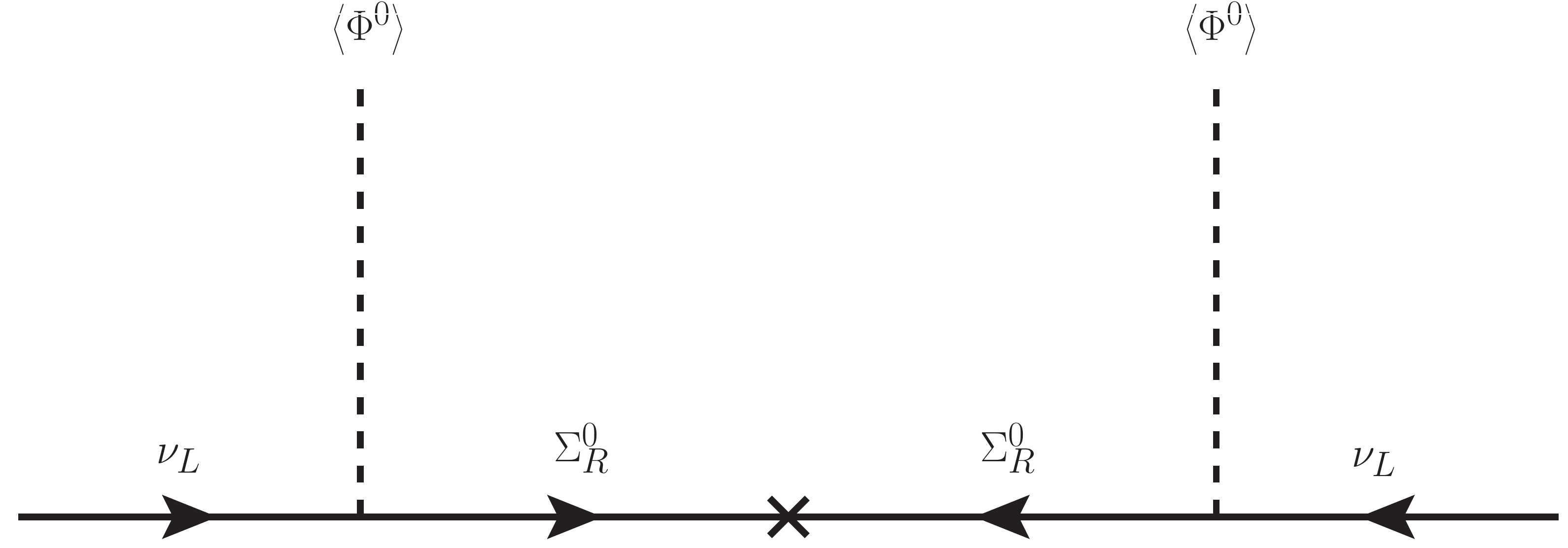}
    \caption{Feynman diagram generating Majorana masses in Type I, II and III seesaw mechanism.}
    \label{MT1}
  \end{figure}

In order to set the stage for our discussion, we start by recalling
the basic features of the operator in \eqref{weinberg}.
There are three different ways of contracting the relevant fields,
associated to the tree-level exchange of different messengers.
The UV completions of these three different contraction possibilities
correspond to the so-called type I, II and III seesaw mechanism:
\be
\underbrace{\underbrace{\bar{L}^c \otimes \Phi}_1 \otimes \underbrace{\Phi \otimes L}_{1}}_{ \textnormal{Type I }} \hspace{0.2cm} , \hspace{0.5cm}
\underbrace{\underbrace{\bar{L}^c \otimes L}_3 \otimes \underbrace{\Phi \otimes \Phi}_{3}}_{\textnormal{Type II }} \hspace{0.2cm} , \hspace{0.5cm}
\underbrace{\underbrace{\bar{L}^c \otimes \Phi}_3 \otimes \underbrace{\Phi \otimes L}_{3}}_{\textnormal{Type III }} \hspace{0.2cm}
\label{wopcont}
\ee
In \eqref{wopcont} the underbrace denotes a $SU(2)_L$ contraction of
the fields involved. The number under the brace denotes the $SU(2)_L$
transformation of the contracted fields. Although not explicitly
written, the global contraction should always be a singlet for the
operator to be allowed by $SU(2)_L$ gauge symmetry. The UV complete
realization of these contractions results into the three well known
seesaw variants as shown in a diagramatical way in Figure \ref{MT1}.
 
  Figure \ref{MT1} illustrates three UV--complete seesaw realizations
  of the same Weinberg operator of \eqref{wopcont}, differing from
  each other in the nature of the messenger fields involved.  
  In the left--panel diagram of Fig.~\ref{MT1}, corresponding to
  type-I seesaw, the field $N_R$ is a heavy fermion which transforms
  as a singlet under $SU(2)_L$ and carries no $SU(3)_C$ or $U(1)_Y$
  charge.
 In the middle  diagram corresponding to type-II seesaw, the field
  $\Delta^0$ is the neutral component of a heavy scalar multiplet
  transforming as triplet under $SU(2)_L$. 
  In the right--panel diagram corresponding to type-III seesaw,
  $\Sigma_R^0$ is the neutral component of the heavy fermion multiplet
  transforming as triplet under $SU(2)_L$ symmetry.

  Notice that the three possible messenger fields and their $SU(2)_L$
  transformation properties arise from the different possibilities of
  field contractions of the Weinberg operator as shown in
  \eqref{wopcont}.
  The possibility where both $\bar{L}^c \otimes L$ and
  $\Phi \otimes \Phi$ contract to a singlet is forbidden, since
  $\Phi \otimes \Phi$ is symmetric, while the singlet contraction is
  antisymmetric, and therefore vanishes. Even in the presence of
  another Higgs doublet, the singlet contraction would vanish due to
  electric charge conservation~\footnote{Such messengers may, however,
    be used in radiative schemes of neutrino mass
    generation~\cite{zee:1980ai}.}

  Weinberg's dimension-5 operator is the lowest one which can generate
  Majorana neutrino masses. One can easily generalize the above
  discussion to higher dimensional operators (a similar discussion for
  radiative mass generation was given in~\cite{Cepedello:2017eqf}).  In general, using only \sm
  fields, i.e.  a single Higgs doublet $\Phi$, it is easy to show that
  the $SU(2)_L$ symmetry implies that Majorana masses can only be
  generated by odd-dimensional operators. This means that they can
  only arise from operators involving even number of Higgs
  doublets. The general operator allowed by \SM symmetry is
\be
 \frac{1}{\Lambda^{2n+1}}\,  \bar{L}^c \Phi^2 \left(\Phi^\dagger \Phi\right)^n L , \hspace{0.2cm} n \in \{0,1,2,3...\}
\ee 
where $\Lambda$ is the cutoff scale, i.e. the scale at which the new
physics associated to the messengers required for UV completion comes
into play.
 

\subsection{ Models with new Higgs fields}
\label{gweinberg}


Weinberg's operator and its higher dimensional siblings only involve
\sm fields. However, the new physics responsible for generating
neutrino mass might be such that the operators involving only \sm
Higgs are forbidden. Such a scenario can also occur in many contexts.
One well known example is provided by models where supersymmetry is
the origin of neutrino mass~\cite{hirsch:2004he} through the
spontaneous breaking of
R-parity~\cite{masiero:1990uj,romao:1992vu,romao:1997xf}. In this case
the messengers are supersymmetric states.
As an alternative example consider the possibility of global Lepton
number ($U(1)_L$) symmetry, broken to its $Z_n$ subgroups
\cite{Hirsch:2017col}.  Where $U(1)_L$ breaks to a $Z_2$ subgroup, the
Weinberg operator is in principle allowed.
However, in scenarios where $U(1)_L$ breaks to higher
$Z_n$ symmetries then Weinberg's operator can be easily forbidden
\cite{Chulia:2016ngi,Chulia:2016giq,CentellesChulia:2017koy}.  
As a simple example, consider the scenario where $U(1)_L$ is
explicitly broken to a $Z_3$ subgroup by messenger field mass
terms. 
In such a case for the lepton doublet transforming non-trivially under
$Z_3$ say $L \sim \omega$, with $\omega^3 = 1$ and the \sm Higgs being
neutral under $Z_3$ it is easy to see that the Weinberg operator as
well as other higher dimensional ones involving only \sm fields are
all forbidden. 
One can still generate Majorana masses for neutrinos if the new
physics involved also contains additional scalars (carrying nontrivial
$U(1)_L$ or $Z_n$ charges) beyond \sm Higgs. 
In this section we briefly discuss such possibilities, most of which
to the best of our knowledge, have not yet been explored in the
literature.

These new possibilities arise in the presence of new scalars carrying
a nonzero vacuum expectation value (vev), such as a scalar
$\chi \sim 1$, singlet under the $SU(2)_L$ symmetry, or the field
$\Delta \sim 3$, triplet under $SU(2)_L$~\footnote{Here we restrict to
  singlets and triplets, although in principle one can also consider
  scalar fields in higher $SU(2)_L$ multiplets.}.
In order to generate Majorana masses some (or all) of these new
vev-carrying scalars must also be charged under the symmetry that
forbids the Weinberg operator, such as $Z_3$ in the simple example
considered above.

\begin{figure}[!h]
 \centering
    \includegraphics[scale=0.25]{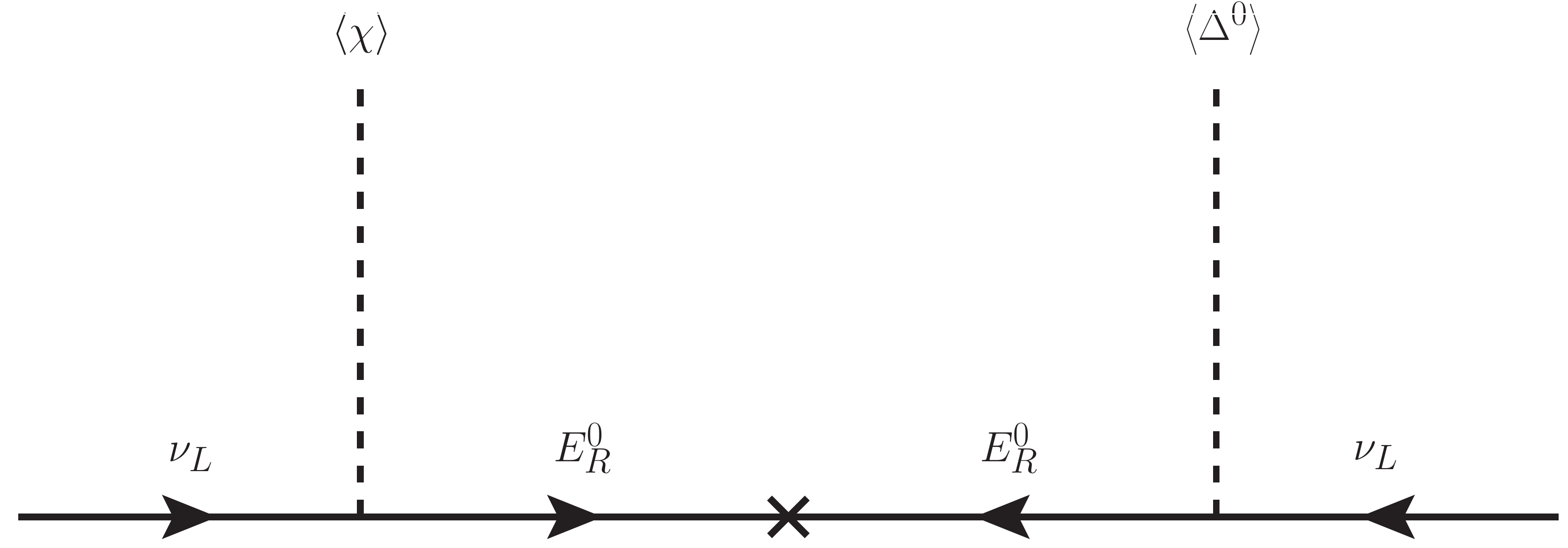}~~~~~~
\includegraphics[scale=0.25]{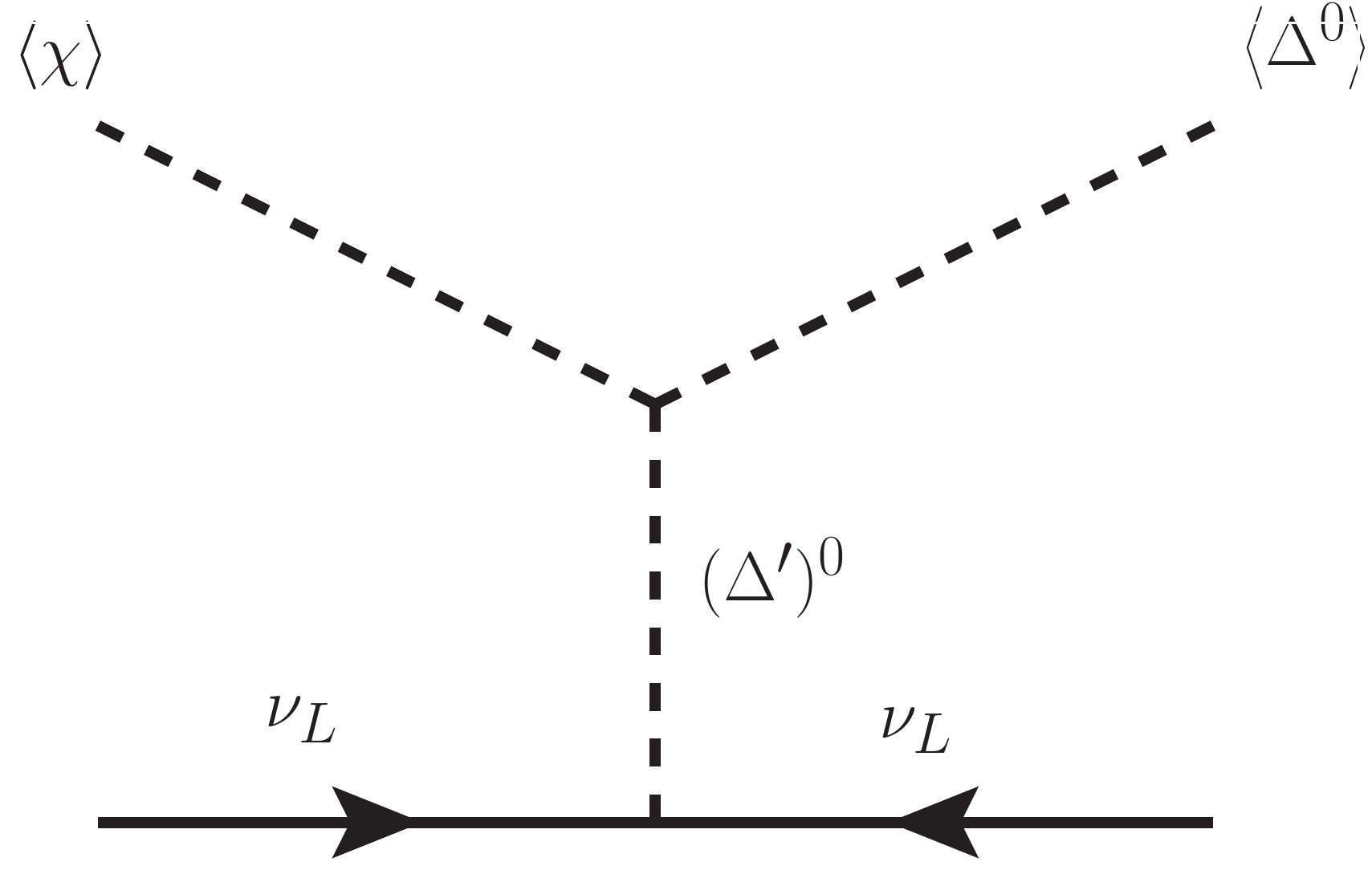}
\caption{ The Feynman diagrams for the two seesaw realizations for the
  generalized Weinberg operator in~\eqref{chi-del-maj}.}
    \label{Mnew}
  \end{figure}

The tree level coupling between $\bar{L} ^c \otimes L$ (dimension 3)
is forbidden by $U(1)_Y$. For the same reason, the dimension-4
operator $\bar{L} ^c \otimes \chi \otimes L$ is also forbidden, unless
$\chi$ is charged under $U(1)_Y$.  In such case electric charge
conservation prevents it from having a vev. However, one could in
principle write the dimension 4 operator with a triplet~\footnote{Note
  that higher $SU(2)_L$ scalar multiplets can not couple to $L$ at the
  dimension 4 level, due to $SU(2)_L$ symmetry.}:
\be
\bar{L}^c \Delta L.
\ee
Though in principle allowed, this would require a tiny vev for the
triplet $\Delta$, in order to account for the smallness of neutrino
masses.
An appealing possibility is to have this term forbidden by symmetry,
and to consider instead, ``Weinberg like'' dimension 5 operators of
the form
\be
\frac{1}{\Lambda} \, \bar{L}^c \otimes X \otimes Y \otimes L
\label{weinberg-gen}
\ee
involving two different~\footnote{Note that the same symmetry
  forbidding the dimension 4 term may also forbid the conventional
  dimension 5 Weinberg operator.} color singlet scalar fields $X$ and
$Y$, transforming as multiplets of $SU(2)_L$ and with appropriate
$U(1)_Y$ charges so as to make \eqref{weinberg-gen} \SM invariant.
Of course, if we take $X \equiv Y \equiv \Phi$, such generalized
Weinberg operator in \eqref{weinberg-gen} reduces to the standard one
in \eqref{weinberg}.
Note that the simplest possibility for generalized Weinberg operator
where $X = Y = \chi$ where $\chi$ is a $SU(2)_L$ singlet, is forbidden by
charge conservation.
However, in general the $X$ and $Y$ fields can be distinct
and this  opens up more possibilities. 
In particular, if $X$ transforms as the ``$n$-th multiplet'' under
$SU(2)_L$, then $Y$ has to transform as $n-2$, $n$ or $n+2$ multiplet
of $SU(2)_L$ with appropriate $U(1)_Y$ charges. 
For example, if the $X$ field is a $SU(2)_L$ singlet, $X \equiv \chi$,
then the $Y$ field must be either a $SU(2)_L$ singlet or
triplet. Therefore, the first operator allowed by gauge invariance
would be
$ \frac{1}{\Lambda} \,\bar{L}^c \otimes \chi \otimes \Delta \otimes
L$.  This operator can be opened up:
\be 
\underbrace{\underbrace{\bar{L}^c \otimes \chi}_2 \otimes
  \underbrace{\Delta \otimes L}_{2}}_{ \textnormal{Type I like}}
\hspace{0.2cm} , \hspace{0.5cm} \underbrace{\underbrace{\bar{L}^c
    \otimes L}_3 \otimes \underbrace{\chi \otimes
    \Delta}_{3}}_{\textnormal{Type II like}}~,
    \label{chi-del-maj}
\ee 
As before, the underbrace denotes the relevant field contraction, and
the number under the brace denotes the $n$-plet transformation
of the contracted fields under $SU(2)_L$. 
In order to preserve the gauge invariance, the global contraction
should be an $SU(2)_L$ singlet.
The first type-I like possibility in \eqref{chi-del-maj} requires
vector-like heavy leptons $E_L, E_R$ transforming as $SU(2)_L$
doublet, as shown in the left  diagram in Fig.~\ref{Mnew}.
The second, type-II like, requires the presence of an additional
$SU(2)_L$ triplet scalar $\Delta'$ carrying an induced vev through its
coupling with the $\chi, \Delta$ fields, see the right diagram in
Fig.~\ref{Mnew}.

  There are other possible choices for the $X$ and $Y$ scalars. For
  example, in case of $X \equiv \Phi$ i.e. a $SU(2)_L$ doublet, the
  $Y$ field can only transform as a doublet or as a quadruplet ($\Xi$)
  of $SU(2)_L$.  In this case, in addition to the standard Weinberg
  operator
  $\frac{1}{\Lambda} \bar{L}^c \otimes \Phi \otimes \Phi \otimes L $
  we can also have the generalized one
  $\frac{1}{\Lambda} \bar{L}^c \otimes \Phi \otimes \Xi \otimes L $.
  On the other hand, if $X \equiv \Delta$ i.e. a $SU(2)_L$ triplet,
  the $Y$ field can only transform as singlet, triplet or quintuplet
  ($\Omega$) under $SU(2)_L$. 
  In this case, in addition to
  $\frac{1}{\Lambda} \bar{L}^c \otimes \Delta \otimes \chi \otimes L $,
  already discussed, one has two other generalized Weinberg operators
  $\frac{1}{\Lambda} \bar{L}^c \otimes \Delta \otimes \Delta \otimes L
  $
  as well as
  $\frac{1}{\Lambda} \bar{L}^c \otimes \Delta \otimes \Omega \otimes L
  $.
  These latter cases seem not to have been explored in the literature.
  However, we will not develop the detailed analysis of the new fields
  required for their UV--completion in this paper.

  Note also that the dimension 5 operator of \eqref{weinberg-gen} can
  easily be generalized to other higher dimensional operators as
\be
\frac{1}{\Lambda^{n-1}} \, \bar{L}^c \otimes X_1 \otimes \cdots \otimes X_n \otimes L
\ee
where $X_i$; $i = 1, \cdots , n$ are scalar fields, at least some of
which lie in different $SU(2)_L$ multiplets, and carry appropriate
$U(1)_Y$ charges.
Again, we will not explore such possibilities any further.


\section{Dirac neutrinos}
\label{sec:dirac-neutrinos}

With the above brief recap of the familiar case of Majorana neutrinos,
we now move on to the possibility of naturally small Dirac neutrino
masses arising from dimension 5 generalized ``Weinberg-type''
operators.
Indeed, there has been a renewed interest in Dirac neutrino mass
generation mechanisms in the recent literature~\cite{Abbas:2013uqh,
  Ma:2014qra, Okada:2014vla, Ma:2015rxx, Ma:2015raa, Ma:2015mjd,
  Chen:2015jta, Valle:2016kyz, Bonilla:2016zef, Bonilla:2016diq,
  Chulia:2016ngi, Chulia:2016giq, Reig:2016ewy, Ma:2016mwh,
  Wang:2016lve, Borah:2016zbd, Addazi:2016xuh, Abbas:2016qbl, Heeck:2013rpa,
  Yao:2017vtm, Ma:2017kgb,CentellesChulia:2017koy, Borah:2017leo,
  Wang:2017mcy, Bonilla:2017ekt, Hirsch:2017col,Srivastava:2017sno,
  Borah:2017dmk,Yao:2018ekp}.
Some of these works have also considered several possibilities of
generation of neutrino masses using operator methods like ours
\cite{Ma:2016mwh, Wang:2016lve, Yao:2017vtm, Yao:2018ekp}. However, a systematic
classification of all possibilities at a given dimension for
generalized Weinberg operators is still lacking. In this paper and a
follow up work we aim to consider all possible generalized Weinberg
operators at dimension 5 and dimension 6 level. Since there are
actually infinite such possibilities, we will restrict our discussion
only to operators involving scalars which transform either as singlet,
doublet or triplet under $SU(2)_L$ symmetry. Furthermore we will only
consider the seesaw high energy completions of these operators and
will not consider the various loop realizations of these
operators. Our method illustrated here can easily be generalized to
include higher $SU(2)_L$ multiplets as well as loop realizations of
the operators. 
  
Before beginning our discussion on possible Dirac seesaw completions
of generalized Weinberg operators, let us mention certain generic
conditions which must be satisfied in order to have Dirac neutrinos.
By definition, a Dirac fermion can be viewed as two chiral fermions,
one left handed and other right handed, having exactly degenerate
masses~\cite{Schechter:1980gr}. 
Thus in order to have massive Dirac neutrinos one must extend the \sm
particle content by adding the right handed partners of the
known neutrinos, $\nu_R$, being singlets under \SM gauge
symmetry.
Second, owing to the color and electric charge neutrality of
neutrinos, an additional exactly conserved symmetry beyond the \sm
gauge symmetries is required to protect the Diracness of neutrinos.
Several ways have been proposed to protect the Dirac nature of
neutrinos, involving symmetries such as an extra $U(1)$ lepton number
symmetry \cite{Ma:2014qra,Ma:2015mjd} or its discrete $Z_n$
subgroups~\cite{Bonilla:2016zef, Bonilla:2016diq}.
In our discussion here we will take lepton quarticity, a discrete
$Z_4$ lepton number symmetry, as a benchmark symmetry used to protect
the Dirac nature of
neutrinos~\cite{Chulia:2016ngi,Chulia:2016giq,CentellesChulia:2017koy}.
As we will see in Sec.~\ref{sec:DM sector}, such symmetry has an
additional virtue of ensuring the stability of dark matter and is
thus a very attractive possibility.
  

\subsection{Only Standard Model fields}
\label{dweinberg}


Here we consider the possibility of having just the minimal \sm Higgs
doublet augmented by three right handed singlet neutrinos. 
The presence of a lepton quarticity symmetry under which both the
lepton doublet $L$ and the right handed neutrinos $\nu_R$ transform as
$z$, with $z^4 = 1$, ensures that the neutrinos remain Dirac particles
provided this symmetry remains exact~\cite{Chulia:2016ngi,
  Chulia:2016giq, CentellesChulia:2017koy}.
This implies that all scalar fields which develop vev should be
neutral under the $Z_4$ quarticity symmetry.
This requirement has profound implications for the stability of dark
matter, as we will discuss in section \ref{sec:DM sector}.  With
only \sm Higgs the simplest possibility to generate Dirac mass is
through the dimension 4 Yukawa interaction
\begin{equation}
y_\nu \,  \bar{L} \Phi^c \nu_R
 \label{nyuk}
\end{equation}
where $y_\nu$ is the Yukawa coupling constant.
Although this possibility is allowed on theory grounds, it leaves the
smallness of neutrino masses unexplained, implying the need for a tiny
Yukawa coupling for neutrinos ($ y_\nu \sim \mathcal{O}(10^{-13})$).
A more attractive possibility would be to forbid this term by an
adequate symmetry and to obtain naturally small neutrino masses
through generalized Weinberg operators or their higher dimensional
counterparts. 
Such a scenario can easily arise in many ways, for example, from a
simple $Z_2$ symmetry \cite{Chulia:2016ngi}, from flavor symmetries
\cite{Chulia:2016giq, CentellesChulia:2017koy} or from an
unconventional $U(1)_{B-L}$ symmetry~\cite{Ma:2014qra,Ma:2015mjd}.

In general, using only the \sm Higgs doublet $\Phi$, the \SM
symmetry implies that the only allowed dimensions for the operators
that can induce Dirac neutrino masses are even, i.e.  operators
involving odd number of Higgs doublets, namely
\be
\frac{1}{\Lambda^{2n}} \bar{L} \Phi^c \left(\Phi^\dagger \Phi \right)^n \nu_R , \hspace{0.2cm} n \in \{0,1,2,3,4...\}
\ee 
Therefore, after the dimension 4 Yukawa term of \eqref{nyuk}, the next
allowed operator involving only \sm Higgs would be of dimension
6. 
Here we restrict ourselves to the discussion only of dimension 5
operators, and hence we will not develop this possibility and its
various UV--completions in detail. 


\subsection{Models with new Higgs fields}
\label{dgweinberg}


As argued before, at the dimension 5 level, Dirac neutrino masses can
only arise from generalized Weinberg operators involving additional
scalars beyond the \sm Higgs. 
As before, we will restrict our discussion to generalized Weinberg
operators involving singlet ($\chi$), doublet ($\Phi$) and triplet
($\Delta$) fields only.
However, our analysis can easily be generalized to the cases of
scalars transforming as higher $SU(2)_L$ multiplets. 
The generalized dimension 5 Weinberg operator for Dirac neutrinos is
given by
\be
\frac{1}{\Lambda} \, \bar{L} \otimes X \otimes Y \otimes \nu_R
\label{dgenw}
\ee
where $X$ and $Y$ are scalar fields transforming as some $n$-plets of
$SU(2)_L$ with appropriate $U(1)_Y$ charges. 

Invariance of \eqref{dgenw} under $SU(2)_L$ symmetry implies that, if
$X$ transforms as a $n$-plet under $SU(2)_L$, then $Y$ must transform
either as a $n+1$-plet, or a $n-1$-plet under $SU(2)_L$ symmetry. 
For example, if we take $X$ to be a singlet then $Y$ should be a
doublet. If we take $X$ to be a doublet then $Y$ can only be a singlet
(equivalent to the previous case) or a triplet. 

\begin{figure}[H]
 \centering
  \includegraphics[scale=0.18]{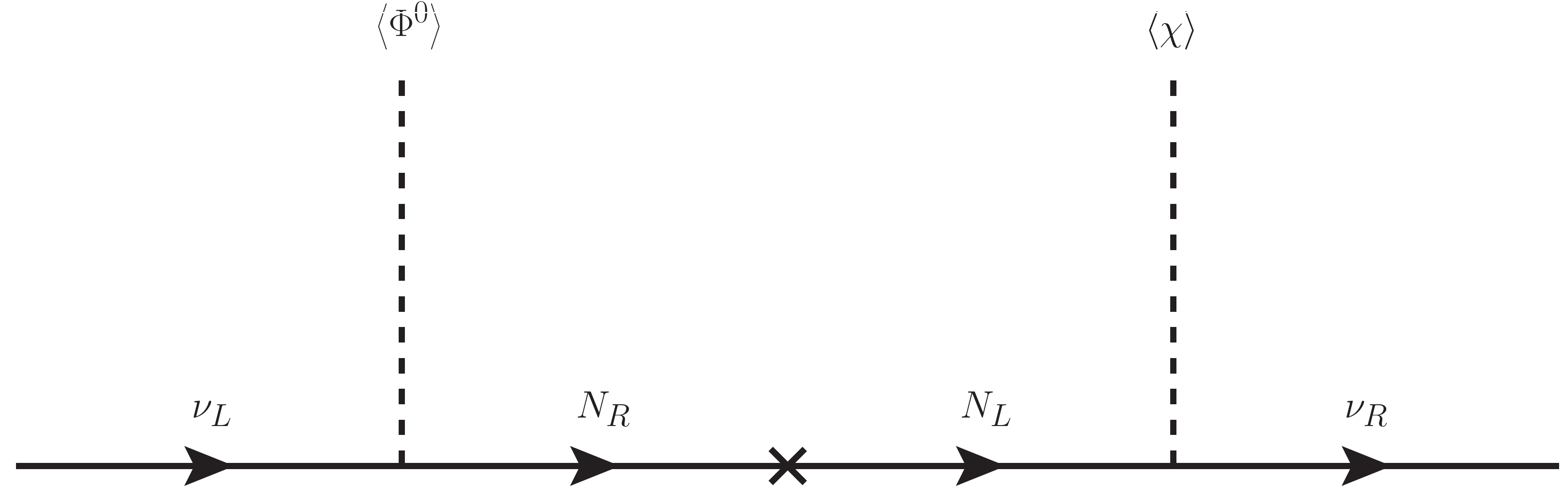}~~~~~~
 \includegraphics[scale=0.18]{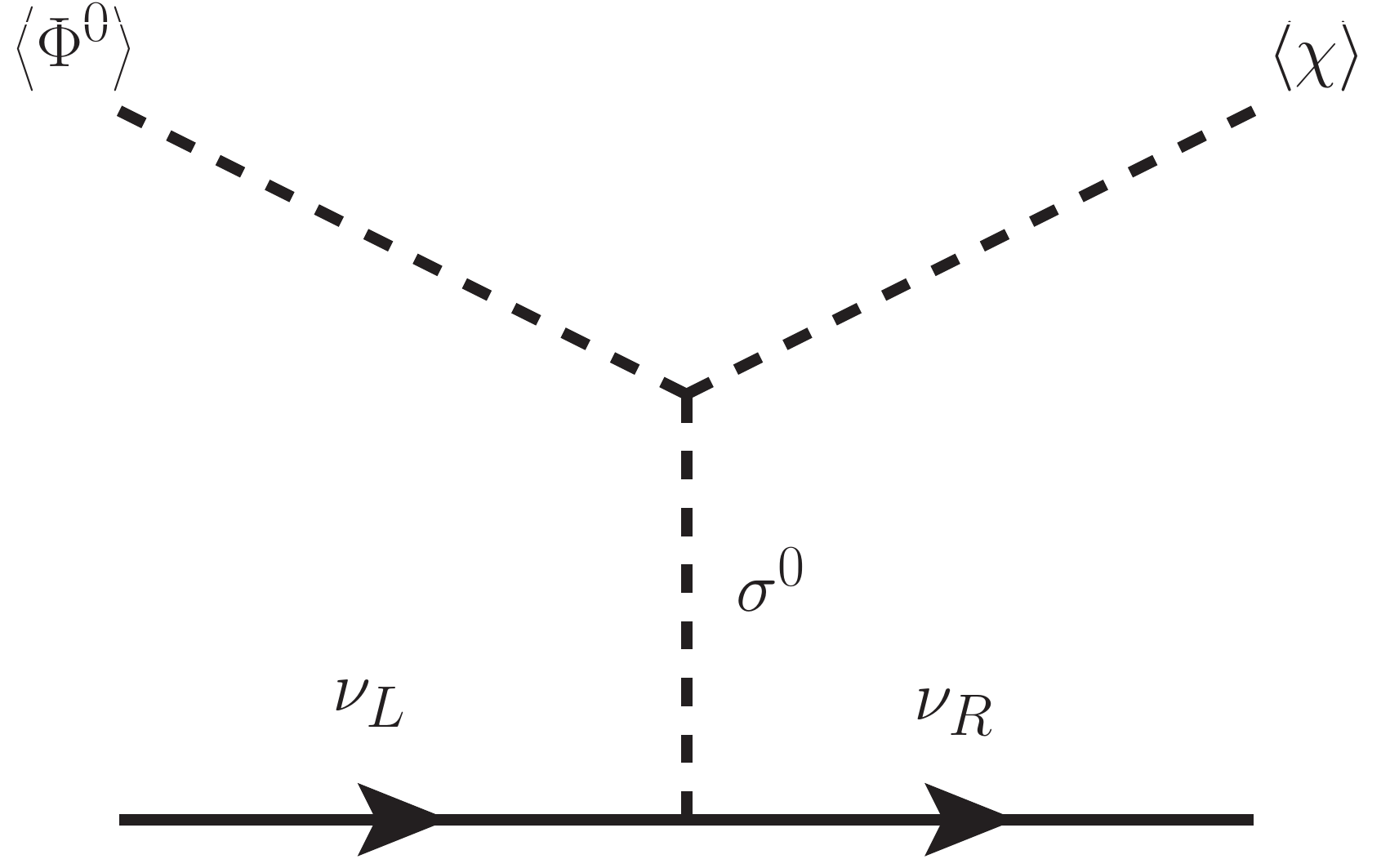}~~~~~~
 \includegraphics[scale=0.18]{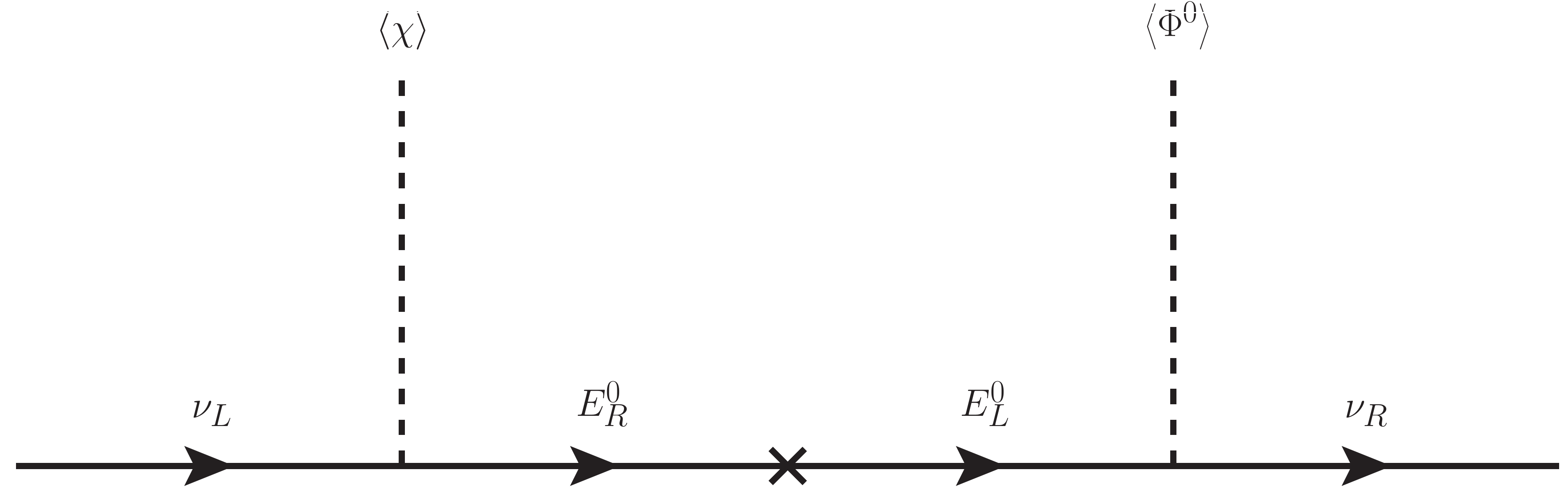}
    \caption{Feynman diagrams representing the Dirac Type I, II and III seesaw analogues.}
    \label{D1}
  \end{figure}

Each of these cases leads to different $SU(2)_L$ contractions which,
as we will see shortly, will lead to different seesaw UV completions.
For example, for the case $X = \chi$ i.e. a singlet under $SU(2)_L$
symmetry and $Y = \Phi$ i.e. a doublet under $SU(2)_L$ symmetry, we
have the following possible contractions, which can be viewed as Dirac
analogues of the type I, II and III Majorana seesaw mechanism,
\be
\underbrace{\underbrace{\bar{L} \otimes \Phi^c}_1 \otimes \underbrace{\chi \otimes \nu_R}_{1}}_{\textnormal{Type I analogue}} \hspace{0.2cm} , \hspace{0.5cm}
\underbrace{\underbrace{\bar{L} \otimes \nu_R}_2 \otimes \underbrace{\Phi^c \otimes \chi}_{2}}_{\textnormal{Type II analogue}} \hspace{0.2cm} , \hspace{0.5cm}
\underbrace{\underbrace{\bar{L} \otimes \chi}_2 \otimes \underbrace{\Phi^c \otimes \nu_R}_{2}}_{\textnormal{Type III analogue}} \hspace{0.2cm}, 
\ee
where again the underbrace denotes a contraction of the fields
involved and the number under the brace denotes the $n$-plet
contraction of $SU(2)_L$ to which the fields contract. Note that invariance under $U(1)_Y$ requires that $\Phi^c$ should appear in this operator. The global contraction should be a singlet in order that the operator is allowed
by $SU(2)_L$. The seesaw completion of these operators will lead to
three different possibilities. These diagrams are shown in figure
\ref{D1}.
  As in the previous cases, the fields $N_L$ and $N_R$ are heavy
  fermions transforming as singlets under $SU(2)_L$, $\sigma^0$ is the
  neutral component of a $SU(2)_L$ doublet scalar, and $E_L^0$ and
  $E_R^0$ are the neutral components of the heavy vector like fermions
  transforming as doublets under $SU(2)_L$.
Note that the new $SU(2)_L$ doublet scalar $\sigma$ must be a new
scalar, and cannot be identified with the Standard model Higgs $\Phi$,
as such an identification will also imply presence of a tree-level
Dirac neutrino mass term we assumed to be forbidden by symmetries. 

Some of the corresponding UV--complete theories have been discussed in
the literature, while others have not. For example, explicit models
employing type I Dirac seesaw have already been realized in
\cite{Ma:2014qra, Chulia:2016giq, Chulia:2016ngi,
  CentellesChulia:2017koy}, while explicit models for type II
  Dirac seesaw where considered in \cite{Valle:2016kyz,
    Bonilla:2016zef, Reig:2016ewy, Bonilla:2017ekt}. 
  In contrast, to the best of our knowledge, a full--fledged
  UV--complete theory using the Dirac type III seesaw has so far not
  been explicitly developed.\\[-.2cm] 

\begin{figure}[H]
 \centering
 \includegraphics[scale=0.18]{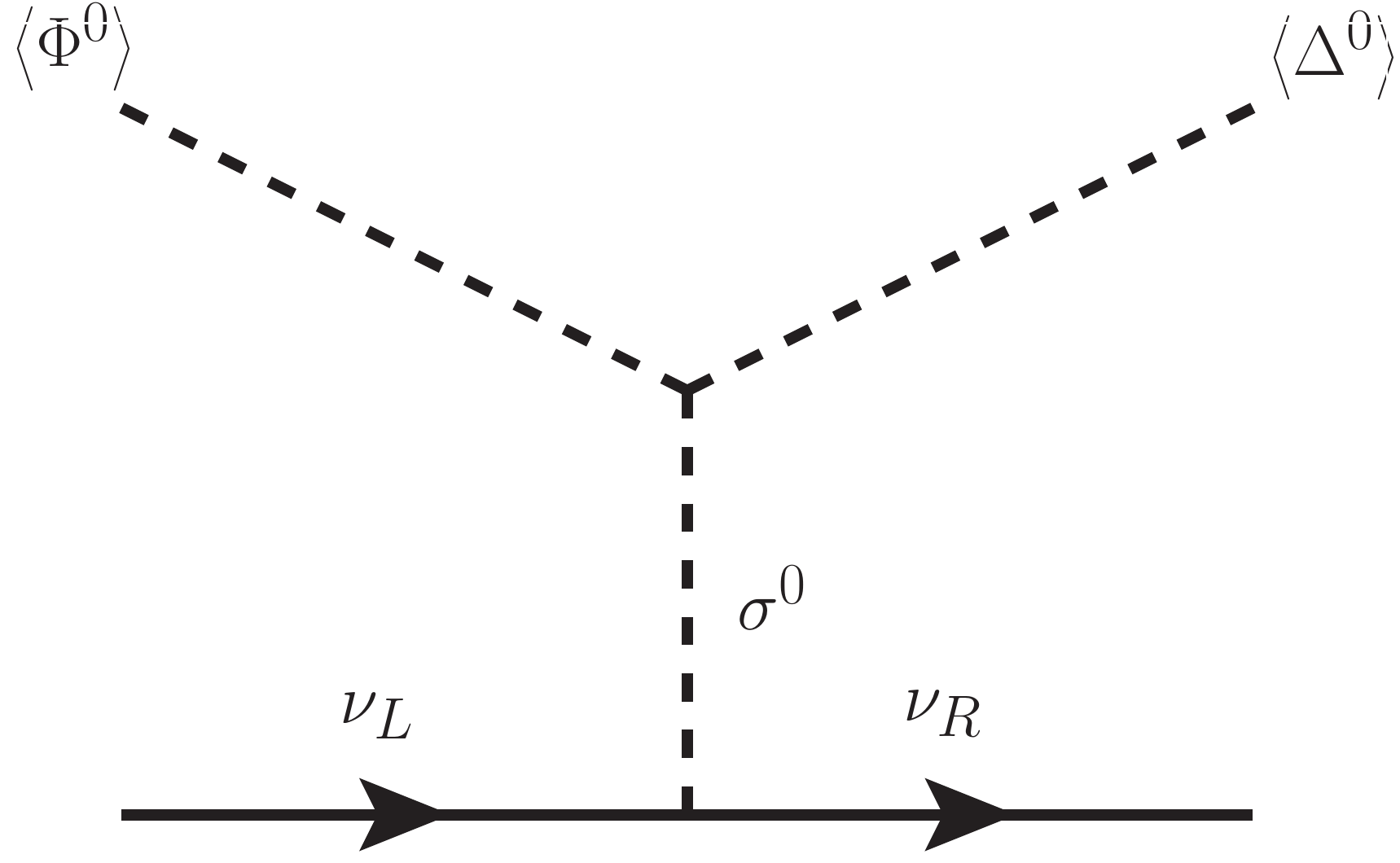}~~~~~~
  \includegraphics[scale=0.18]{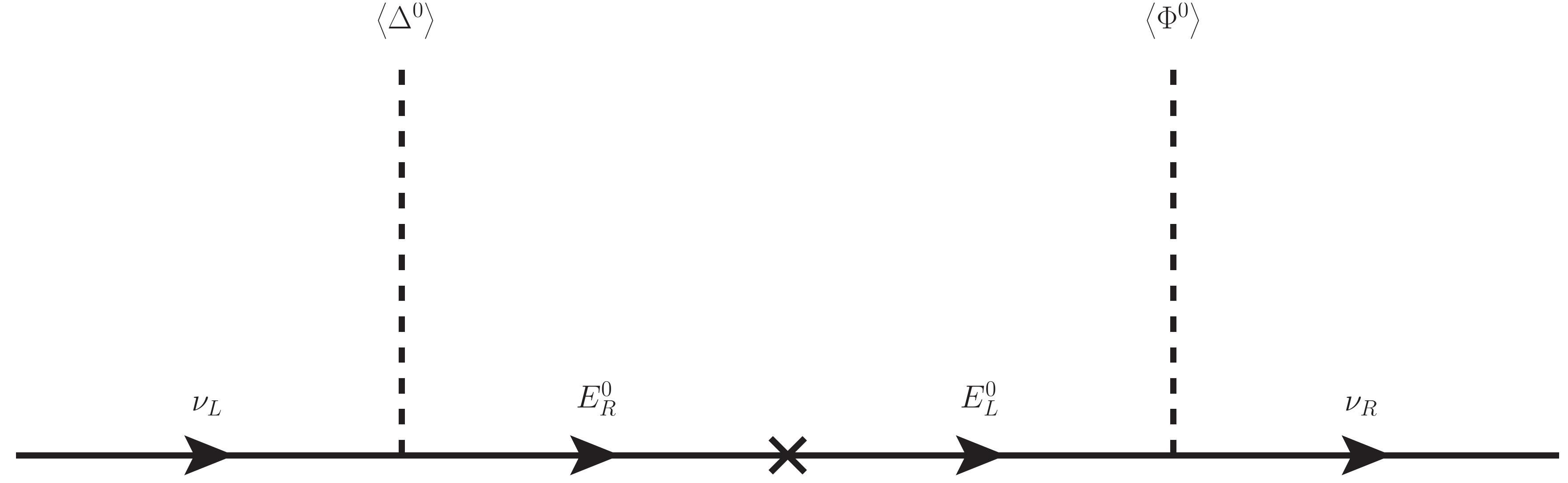}~~~~~~
\includegraphics[scale=0.18]{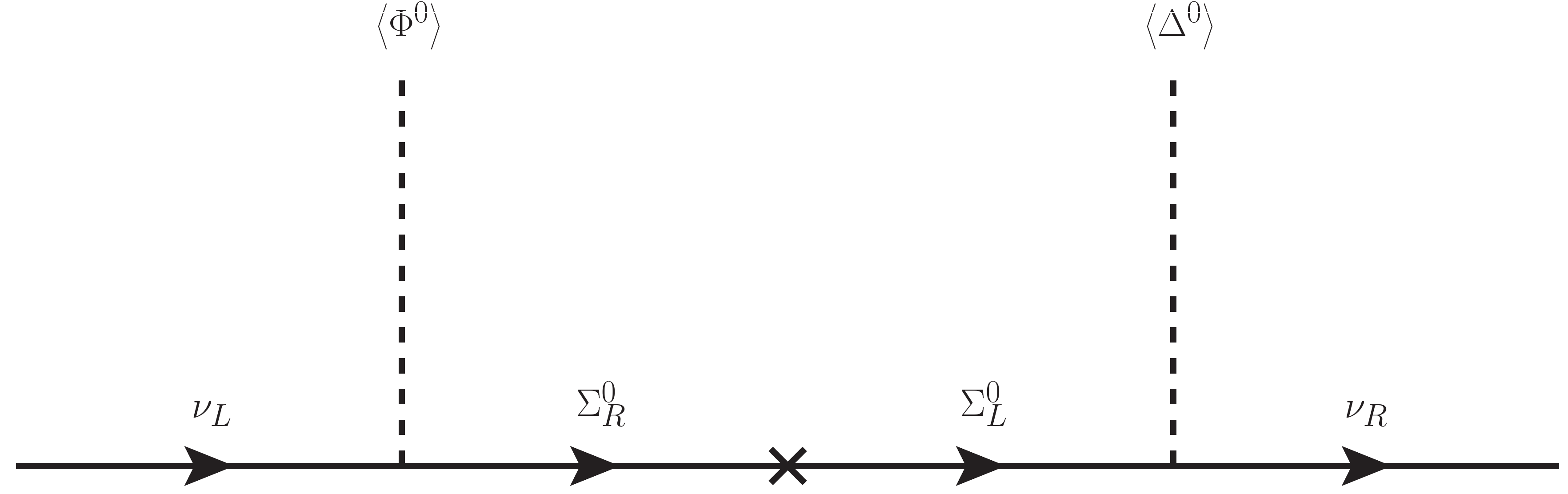}~
\caption{Feynman diagrams for Type-II and Type-III Dirac seesaw mechanism.}
    \label{D2}
  \end{figure}

  Going beyond singlets and doublets opens up still more
  possibilities. For example, taking $X = \Phi$ and $Y = \Delta$
  yields  
\be
\underbrace{\underbrace{\bar{L} \otimes \nu_R}_2 \otimes \underbrace{\Phi \otimes \Delta}_{2}}_{\textnormal{Type II like}} \hspace{0.2cm}, 
\hspace{0.5cm}
\underbrace{\underbrace{\bar{L} \otimes \Delta}_2 \otimes \underbrace{\Phi \otimes \nu_R}_{2}}_{\textnormal{Type III like}} \hspace{0.2cm} , \hspace{0.5cm}
\underbrace{\underbrace{\bar{L} \otimes \Phi}_3 \otimes \underbrace{\Delta \otimes \nu_R}_{3}}_{\textnormal{Type III like}} \hspace{0.2cm}
\label{dtrip}
\ee
The underbrace denotes, as before, field contraction, and the number
under the brace denotes the $n$-plet contraction of $SU(2)_L$ to which
the fields reduce.
The global contraction should be an $SU(2)_L$ singlet. Note that for this operator we have two possibilities for $U(1)_Y$ charge of $\Delta$.
Apart from the operator in \eqref{dtrip} (which has $\Delta$ with $U(1)_Y = -2$) another operator namely $\bar{L} \Phi^c \Delta_0 \nu_R$ with $\Delta_0$ carrying $U(1)_Y = 0$ is also possible. The diagrams for this case will be identical to those discussed here but the hypercharges of the intermediate fields will be different.
  Note that one can always induce the vevs of either $\chi$ or
  $\Delta$ with the coupling to a pair of $\Phi$'s. Such operators will
  have dimension 6 and will be discussed in a follow--up work. The
  diagrams leading to the seesaw completion of \eqref{dtrip} are
  shown in Figure \ref{D2}.
  where, as before, the field $\Delta^0$ is the neutral component of
  the $SU(2)_L$ triplet, $\Sigma_L^0$ and $\Sigma_R^0$ are the neutral
  components of the heavy $SU(2)_L$ triplet fermions, $\sigma^0$ is
  the neutral component of an $SU(2)_L$ doublet scalar $\sigma$, while
  $E_L^0$ and $E_R^0$ are the neutral components of the heavy
  vector--like fermions transforming an $SU(2)_L$ s doublet. 
  As before, owing to the symmetry requirements, $\sigma$ must be a
  new $SU(2)_L$ doublet, distinct from the Standard model Higgs $\Phi$.
  There have been so far no dedicated study of UV--complete theories
  in literature corresponding to these new possibilities.
  
  Going yet to higher multiplets of $SU(2)_L$ will open up novel ways
  to generate Dirac neutrino mass at the dimension 5 level.  These can
  easily be realized following our procedure in a straightforward way,
  so here we skip the details.
  Furthermore, as already discussed, Dirac neutrino masses can also be
  generated at higher dimensions. The general operator for such a
  scenario involves several scalar fields $X_i$; $i = 1, \cdots , n$
  which can be different $SU(2)_L$ multiplets, carrying appropriate
  $U(1)_Y$ charges, as follows
\be
\frac{1}{\Lambda^{n-1}} \, \bar{L} \otimes X_1 \otimes \cdots \otimes X_n \otimes \nu_R
\label{dhwop}
\ee
Note that some of the $X_i$ may coincide. In this paper we will not
develop such possibilities, which we plan to do as a follow up work.

\section{Dirac neutrinos and Dark Matter Stability}
\label{sec:DM sector}

As stated before, if neutrinos are Dirac particles an additional
symmetry is required to protect their Dirac nature.
Various types of additional symmetries have been used in many
different new physics scenarios. 
However one of the simplest possibilities for such a symmetry is a
discrete residual $Z_n$ subgroup arising from the breaking of the
usual \sm global $U(1)_L$ symmetry, by the new physics.
Such discrete lepton number symmetries can indeed ensure that the
Dirac nature of neutrinos can be consistently preserved. 
As an example, the discrete $Z_4$ symmetry, called lepton quarticity,
has been employed in actual UV--complete models in
Refs.~\cite{Chulia:2016ngi,Chulia:2016giq, CentellesChulia:2017koy}.

An attractive feature of such constructions is that the same symmetry
which ensures the Dirac nature of neutrinos can also provide stability
to the dark matter particle, thus linking dark matter stability with
Dirac nature of neutrinos in an intimate way~\cite{Chulia:2016ngi}.

In this section we argue that this connection may be quite deep. The
Dirac nature of neutrinos and dark matter stability can be
accomplished by the same symmetry, irrespective of the details of the
particular mass model, and of the nature of their UV--completion.
For illustration, take the case of $Z_4$ lepton quarticity
symmetry. The $Z_4$ group admits only one--dimensional irreducible
representations, conveniently represented by the fourth roots of unity
$z, z^4 =1$. 
Such quarticity symmetry in context of Dirac neutrinos may arise as a
residual subgroup of $U(1)_L$ or $U(1)_{B-L}$. Under the quarticity
symmetry the lepton doublets $L$ and right handed neutrinos $\nu_R$
transform as $z$.~\footnote{ If such quarticity symmetry is a
  remnant of the $U(1)_{B-L}$ symmetry, then the quarks can also
  transform as $z$ under quarticity.}
Since the quarticity symmetry is preserved, no scalar fields which
obtain a non-zero expectation value should be charged under $Z_4$.
Thus we have:
\begin{eqnarray}
\text{If} \, \, \langle X_i \rangle & \neq & 0,  \, \, \text{then} \,\, X_i \, \sim \, 1 \, \, \text{under} \, \, Z_4\\
\text{If} \, \, \zeta_i &\nsim & 1 \, \, \text{under} \, \, Z_4, \, \, \text{then} \, \, \langle \zeta_i \rangle \, \, = \, \, 0~.
\end{eqnarray}
where $X_i, \zeta_i$; $i = 1, \cdots n$ denote the scalar fields. 
The above constraints have a profound effect in a completely
unexpected direction. Consider for example, the generalized Weinberg
operator for Dirac neutrinos of \eqref{dgenw}.
Here, the scalar fields $X,Y \sim 1$ under $Z_4$ to preserve the
Diracness of neutrinos. \\[-.2cm]

Consider now another scalar field $\zeta$, singlet under \sm gauge
symmetry, but tranforming as $\zeta \sim z$ under the $Z_4$, hence
carrying no vev i.e. $\vev{ \zeta} = 0$.
Its interactions with the other fields are severely restricted by
$Z_4$. Indeed, notice that the Yukawa coupling of $\zeta$ with any
fermion, as well as the cubic couplings with the scalars $X_i$,
i.e.~$X_i^\dagger X_i \zeta$, which would lead to its decay, are all
forbidden by the $Z_4$. 

\begin{figure}[H]
 \centering
\includegraphics[scale=0.49]{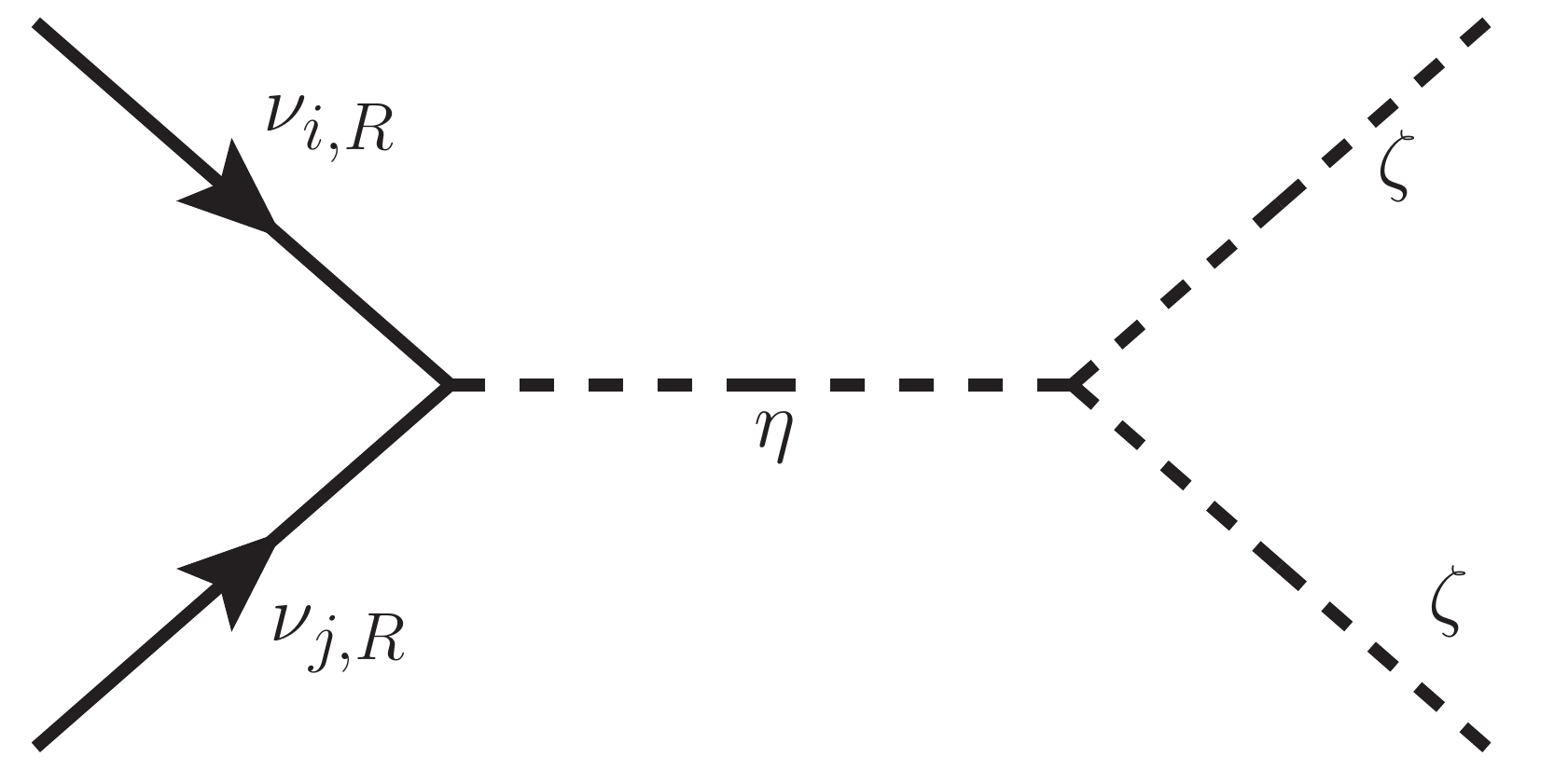}
\caption{Coupling between right handed neutrinos $\nu_R$ and the
  WIMP scalar dark matter particle $\zeta$, mediated by the scalar $\eta$. }
    \label{dm}
  \end{figure}

In order to make sure that the $Z_4$ symmetry stabilizing the $\zeta$
field has indeed a (discrete) lepton number nature, one needs a
``messenger'' field $\eta$ connecting the $Z_4$ charges of neutrinos
with that of $\zeta$~\cite{Chulia:2016ngi}.
This is easily accomplished by having a new scalar field $\eta$,
singlet under the \SM gauge group, but transforming as $\sim z^2$
under $Z_4$ symmetry. 
Since $\eta$ has nontrivial $Z_4$ charge, we must require
$\vev{ \eta} = 0$.
Owing to its $Z_4$ charge, $\eta$ has a Yukawa coupling to right
handed neutrinos, $\bar{\nu}^c_R \nu_R \eta$, as well as a cubic
$\eta \zeta \zeta$ coupling term, as seen in Fig.~\ref{dm}.
Both terms are invariant under the $Z_4$ as well as the \sm gauge
group. The $\eta$ field thus connects neutrinos to dark matter,
through the diagram shown in the Fig.~\ref{dm}.

Notice that the $Z_4$ symmetry remains conserved even after \SM
symmetry breaking takes place. Hence the stability of the $\zeta$
field is also ensured after electroweak symmetry breaking, making
it a {\sl bona fide} dark matter candidate, whose stability emerges
from the same quarticity symmetry responsible for protecting the Dirac
nature of neutrinos.
Due to the presence of extra singlet scalar bosons, its relic density
is somewhat model--dependent, due to new diagrams contributing to the
dark matter annihilation.  
Nevertheless, dark matter relic abundances are typically WIMP-like.

Concerning its interaction with \sm particles, which determine its
nuclear recoil direct detection cross section, it proceeds through the
so-called Higgs portal and has been discussed in
\cite{Chulia:2016ngi}, see Fig.~4.
Recent direct WIMP dark matter searches in LUX \cite{Akerib:2016vxi},
PandaX \cite{Cui:2017nnn} and Xenon1T \cite{Aprile:2017iyp} lead to
constraints on dark matter masses versus coupling strength which are
slightly improved in comparison to analysis of \cite{Chulia:2016ngi}
For the low mass Higgs--portal dark matter region, below half of the
Higgs boson mass, there are somewhat improved constraints on the Higgs
invisible decay width which come from recent searches at the
LHC~\cite{Basalaev:2017cpw}.

  As a final comment we stress that the above argument can also
  be extended to higher dimensional Weinberg operators, such as those
  in \eqref{dhwop}. Hence the connection between the Dirac nature of
  neutrinos and dark matter stability can be made quite generic,
  irrespective of the dimensionality of the operators leading to Dirac
  neutrino mass generation, as well as the details of
  the particle content involved in the UV--completion.

\section{ Summary  and Discussion}
\label{sec:summary-conclusions-}

Here we have described the various ways to induce neutrino masses
through generalized dimension-5 Weinberg operators, both for the case
of Majorana as well as Dirac neutrinos.
In both cases, the need for extra scalar multiplets beyond the
Standard Model Higgs doublet, implies new possible field
contractions. These are absent in the simple case of the Majorana
seesaw mechanism obtained from the UV--completion of the conventional
Weinberg operator.
We have identified many possibilities which have already been employed
within specific models, as well as novel ones.
We have also noticed that, for the case of Dirac neutrinos, the extra
symmetries required for ``Diracness'' can, rather generically, play a
double role in also ensuring the existence of a stable WIMP dark
matter candidate.

As a final remark, we have stressed that the connection with WIMP dark
matter that we have proposed can be made quite general, irrespective
of model realizations.
We stress that our proposed connection between neutrino masses and
dark matter is rather different from most of the existing ones, such
as the ``scotogenic'' approach~\cite{Ma:2006km}, where the WIMP dark
matter particle is a radiative neutrino mass
messenger~\cite{Hirsch:2013ola,Merle:2016scw}, and whose simplest
realization yields Majorana neutrinos, instead of Dirac neutrinos.

\begin{acknowledgments}
 
  Work supported by the Spanish MINECO grants FPA2017-85216-P and
  SEV2014-0398, and also PROMETEOII/2014/084 from Generalitat
  Valenciana.  The Feynman diagrams were drawn using Jaxodraw
  \cite{Binosi:2003yf}. 

\end{acknowledgments}


\bibliographystyle{bib_style_T1}

\providecommand{\url}[1]{\texttt{#1}}
\providecommand{\urlprefix}{URL }
\providecommand{\eprint}[2][]{\url{#2}}

\end{document}